\documentclass[a4paper,12pt]{article}
\usepackage{epsfig}
\date{\today}
\newcommand{\bmat}{\left(\begin{array}}
\newcommand{\emat}{\end{array}\right)}

\def\lsim{\raise0.3ex\hbox{$\;<$\kern-0.75em\raise-1.1ex\hbox{$\sim\;$}}}
\def\gsim{\raise0.3ex\hbox{$\;>$\kern-0.75em\raise-1.1ex\hbox{$\sim\;$}}}

\begin{document}

\renewcommand{\thefootnote}{\fnsymbol{footnote}}

\rightline{IPPP/01/40}
\rightline{DCPT/01/80}
\rightline{FTUAM 01/19}
\rightline{IFT-UAM/CSIC-01-29}
\rightline{October 2001}

\vskip 1cm
\begin{center}
{\bf \LARGE{The Enigma 
of the Dark Matter\\[10mm]}}
{\large{Shaaban Khalil}$^{1,2}$ \small{and}
\large{Carlos Mu\~noz}$^{3,4}$\\[6mm]}
\small{
$^1$ 
IPPP, Physics Department, Durham University, DH1 3LE, Durham,~~U.~K.
\\
$^2$ Ain Shams University, Faculty of Science, Cairo, 11566, Egypt.
\\
$^3$ Departamento de F\'{\i}sica
Te\'orica C-XI, Universidad Aut\'onoma de Madrid,\\[-0.1cm]
Cantoblanco, 28049 Madrid, Spain. \\
$^4$ Instituto de F\'{\i}sica Te\'orica  C-XVI,
Universidad Aut\'onoma de Madrid,\\[-0.1cm]
Cantoblanco, 28049 Madrid, Spain.
\\[15mm]}
\end{center}

\hrule
\vskip 0.3cm
\begin{center}
\small{\bf Abstract}\\[3mm]
\end{center}

\begin{minipage}[h]{14.0cm}

One of the great scientific enigmas still unsolved, the existence
of dark matter, is reviewed.
Simple gravitational arguments imply that most of the mass in the
Universe, at least 90\%, is some (unknown) non-luminous matter. 
Some particle candidates for dark matter are discussed with particular
emphasis on the neutralino, a particle predicted
by the 
supersymmetric extension of the Standard Model of particle physics.
Experiments searching for these relic particles,
carried out by many groups around the world, 
are also discussed.
These experiments are becoming more sensitive every year and in fact
one of the collaborations claims that the first direct evidence for
dark matter has already been observed.
\end{minipage}

\vskip 0.3cm
\hrule
\vskip 2cm

\section{Why do we need dark matter?
}

One of the most evasive and fascinating enigmas in physics is 
the problem of the dark matter in the Universe.
Most astronomers, cosmologists and particle physicists are
convinced that at least 90\% of the mass of the Universe is
due to some non-luminous matter, the so called
`dark matter'. However, although the existence of dark matter was
suggested 68 years ago, still we do not know its composition.

In 1933 the astronomer Fritz Zwicky 
provided evidence that the mass of the luminous matter (stars) 
in the Coma cluster, which consists of about 1000 galaxies, 
was much smaller than its total mass implied by the 
motion of cluster member galaxies.
But, only in the 1970's the existence of dark matter began to be
considered seriously. Its presence in spiral galaxies was
the most plausible explanation for  
the anomalous rotation curves of these galaxies,
as we will discuss in detail in the next subsection. 
In summary, 
the measured rotation velocity of isolated stars or
gas clouds in the outer parts of galaxies was not
as one should expect from the gravitational attraction due to 
the luminous matter. This lead astronomers to assume that
there was dark matter in and around galaxies. 
Although the nature of this dark matter is still unknown,
its hypothetical existence is not so odd if we remember that the discovery 
of Neptune in 1846 by Galle was due to the 
suggestion of Le Verrier on the basis of the irregular motion of
Uranus. 

\subsection{Dark matter in galaxies}

To compute the rotation velocity of stars or hydrogen clouds
located far away from galactic centres 
is easy. One only needs to extrapolate
the Newton's law, which works outstandingly well for nearby
astronomical phenomena, to galactic distances. Let us recall e.g. that  
for an average distance $r$ of a planet from the center 
of the Sun,
Newton's law implies that 
$v^2(r)/{r} = {G M(r)}/{r^2}$,
where $v(r)$ is the average orbital velocity of the planet,
$G=6.67\times 10^{-11}$ m$^3$ kg$^{-1}$ s$^{-2}$ is the Newton's 
constant and $M(r)$ is the total mass inside 
the orbit.
Therefore one obtains
%
\begin{equation}
v(r)= \sqrt{\frac{G\ M(r)}{r}}
\ .
\label{Newton}
\end{equation}
Clearly, $v(r)$ decreases with increasing radius
since $M(r)$ is constant and given by the solar mass 
$M_{\odot}=1.989\times 10^{30}$ kg.
The credibility of this formula can be deduced 
easily. For example, the mean distance for the Earth to the Sun
is $150\times 10^{6}$ km implying $v=30$ km\ s$^{-1}$.
For Neptune whose mean distance is 30 times bigger the velocity
is 5.4 km\ s$^{-1}$. As it is well known both results for the
velocity of the Earth and Neptune are correct.

\begin{figure}[t]
\begin{center}
\begin{tabular}{c}   
\epsfig{file= 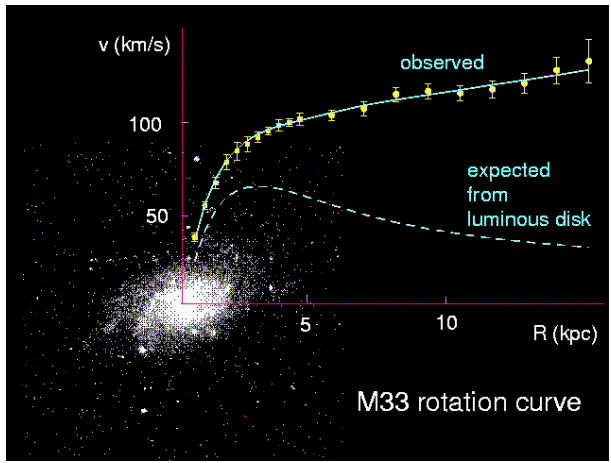, width=8cm}\\
\end{tabular}
\end{center}
\caption{Observed rotation curve of the nearby dwarf spiral galaxy M33,
superimposed on its optical image (from ref.~\cite{Roy}).}
\label{flat}
\end{figure}

In the case of a galaxy, if its
mass distribution can be approximated as
spherical or ellipsoidal, eq.(\ref{Newton}) can also be used
as a estimate.
Thus 
if the mass of the galaxy 
is concentrated in its visible part, one would
expect $v(r) \sim 1/\sqrt{r}$ 
for distances far beyond the visible radius.
Instead, astronomers, by means of the Doppler effect, observe that
the velocity rises towards a constant value 
about 100 to 200 km\ s$^{-1}$.
Thus for large distances $M(r)/r$ is generically constant. 
Hence the mass interior to $r$ increases
linearly with $r$.
An example of this can be
seen in Fig.~\ref{flat}, where the rotation curve of M33, 
one of the about 45 galaxies which form our small cluster, the Local
Group,
is shown. For comparison, the expected velocity from luminous disk
is also shown. Using approximation (\ref{Newton}) and the visible mass
of M33, $4\times 10^{10} M_{\odot}$, one can roughly reproduce
this curve.

This phenomenon has already been observed for about a thousand 
spiral galaxies \cite{Persic},
and in particular also for
our galaxy, the Milky Way.
The most common explanation for these flat rotation curves
is to assume that disk galaxies are immersed in extended dark matter
halos.
While at small distances this dark matter is only a small fraction of
the galaxy mass inside those distances, it becomes a
very large amount at larger distances. 
For instance for\footnote{The parsec (pc) is a unit of distance,
1 pc = 3.26 light-years $\simeq$ 30.8$\times 10^{12}$ km.
For galactic distances, the kiloparsec (kpc),
1 kpc = $10^{3}$ pc, 
is used. On the other hand, for cosmological distances, the megaparsec (Mpc), 
1 Mpc = $10^{6}$ pc, is preferred. For example the Coma cluster
mentioned
above measures about 1.5 Mpc across.} 
$r=10$ kpc 
in Fig.~\ref{flat}, since the observed velocity is
$v\approx 120$ km\ s$^{-1}$
and the expected velocity due to the luminous matter is 
$v_{\mbox{\tiny lum}}\approx 40$ km\ s$^{-1}$, 
one obtains using approximation
(\ref{Newton}) the following ratio between the total mass and
the luminous mass,
$M\approx 9 M_{\mbox{\tiny lum}}$.
Clearly, 
this type of analyses imply that 90\% of the mass in galaxies
is dark.

%

Cosmologists usually express
the present-day mass density averaged over the Universe, $\rho$, in units 
of the so-called critical density,
$\rho_c \approx 10^{-29}$ g cm$^{-3}$,
i.e. they define
$\Omega = \rho/\rho_c$.
We can understand the critical density much like the notion of escape velocity:
$\rho=\rho_c$ ($\Omega =1$) corresponds to the cancellation of kinetic and
(gravitational) potential energies. 
If $\rho > \rho_c$ ($\Omega > 1$) the Universe expands to a maximum,
and then contracts leading to an inverse Big Bang (closed Universe).
If $\rho < \rho_c$ ($\Omega < 1$) the Universe will continue expanding forever 
(open Universe). This is also the situation of the critical case 
$\Omega = 1$, where moreover the geometry of the Universe is flat.
Clearly, the amount of dark matter in the Universe 
will then determine its future 
since $\Omega = \Omega_{\mbox{\tiny lum}} + \Omega_{\mbox{\tiny dark}}$.  

Whereas current observations of luminous matter in galaxies determine
$\Omega_{\mbox{\tiny lum}} \lsim 0.01$, 
analyses of rotation curves imply in fact 
$\Omega \approx 0.1$.
The latter is really a 
lower bound, since 
almost all rotation curves remain flat out to the largest values 
of $r$ where one 
can still find objects orbiting galaxies. We do not really
know how much further the
dark matter halos of these galaxies extend (see e.g. Fig.~\ref{flat}).
The case of the Milky Way,
which has a visible size 
of about 30 kpc across, is 
controversial but might have a halo as large as 200 kpc. 
Since the distance between the Milky Way and M31, the Andromeda galaxy, 
is 350 kpc,
the halos of both galaxies might even touch each other.
Thus we can conclude that galactic rotation curves imply 
\begin{equation}
\Omega \gsim 0.1
\ .
\label{omegagalactic}
\end{equation}

Let us finally remark two points concerning galaxies. First,
although the detection of dark matter is more problematic in
galaxies other than spirals, such as ellipticals,
dwarf irregulars, dwarf spheroidals, lenticulars, etc.,
they also possess important amounts of dark matter \cite{Salucci}.
Second, it is fair to say that a small number of authors suggest
that dark matter is not really necessary to explain rotation
curves. 
Basically their approach consists of modifying the
Newton's law at galactic scales. However, 
these attempts are not only rather ad hoc
in general but also insufficient to account the necessity
of dark matter in scales larger than galactic ones.  
Evidence for dark matter in 
large scale structures is precisely 
the aim of our next discussion.


\subsection{Dark matter in large scales}

The analysis of dark matter in cluster of galaxies becomes more
involved than in galaxies. Different techniques have been used
to compute the value of $\Omega$ \cite{Freedman}. 
For example, the conventional method of studying the motion of
cluster member galaxies seems to point at
values of $\Omega$ larger than those obtained in galactic 
scales. The high temperature of the gas detected in clusters
through its X-ray emission also points at large amounts
of dark matter. 
Finally, the more reliable method of
studying the gravitational lensing confirms the previous conclusions.
Here a cluster acts as a lens which distorts
the light emitted
by quasars and field galaxies in its background, due to the gravitational
bending of light. All these analyses favour a value
\begin{equation}
\Omega \approx 0.2-0.3
\ .
\label{omegacluster}
\end{equation}

Moreover, measurements of velocity flows at scales 
larger than that of clusters
favour also very large amounts of dark 
matter \cite{Freedman}:
%
\begin{equation}
\Omega > 0.3
\ .
\label{omegaflow}
\end{equation}

In fact, theoretical arguments prefer a 
$\Omega = 1$ flat Universe \cite{Guth}, and therefore
even larger amounts of dark matter would be necessary (unless we
take seriously the speculative idea of a non-vanishing 
vacuum energy density $\rho_{\mbox{\tiny c.c.}}$ contribution
to the density of the Universe, i.e. a cosmological constant
contribution $\Omega = \Omega_{\mbox{\tiny lum}} + \Omega_{\mbox{\tiny dark}} 
+ \Omega_{\mbox{\tiny c.c.}}$.)

\subsection{What is dark matter made of?}

Nowadays there is overwhelming observational evidence
for the presence of dark matter. It not only clusters  
with stellar matter forming the galactic halos, but also
exists as a background density over the entire Universe.
Thus the problem is no longer to explain the rotation curves
but to decipher the nature of the dark matter.
As we will explain in detail below, the search of its solution provides
a potentially important
interaction between particle physics and cosmology since
only elementary particles are reliable candidates for the dark matter
in the Universe.
The reason being that they are the only candidates which can be 
present in the right amount to explain the observed density
of the Universe. In addition, they are
necessary to explain structure formation. 
Clumps of neutral particles arose first through gravitational
attraction
and later, when the neutral atoms were formed, these were
gravitationally
attracted by the dark matter to form the galaxies.

Here we will review first possible candidates for dark matter.
In particular we will learn that 
baryonic objects, i.e.
formed from protons and neutrons, such as e.g. gas, brown dwarfs, etc.,
can be components of the dark matter, but more candidates are needed.
Fortunately, particle physics offers various candidates for
dark matter.
Although the 
current Standard Model of particles and interactions \cite{Kalmus} 
does not 
have non-baryonic 
particles that can account for the dark matter,
several extensions of this model do have.
Indeed, detecting non-baryonic dark matter
in the Universe would be a signal for new physics beyond the Standard Model.

We will see that very interesting and plausible candidates for dark
matter are weakly-interacting massive particles (WIMPs).
Long-lived or stable WIMPs can remain from the earliest moments
of the Universe in sufficient number to account for a significant
fraction of  relic dark matter density.
This raises the hope of detecting relic WIMPs directly
by observing their elastic scattering on target nuclei though nuclear
recoils. In fact, recent 
results from a dark matter experiment
of this type at an underground laboratory suggest that the
first direct evidence for dark matter has already been observed.
Although this could be a great discovery, it is fair to say that
this experiment is controversial, mainly because 
another one reported negative search results.

In any case, due to these and other projected experiments, it seems
very plausible that the dark matter will be found in the near future.
Assuming that it is a WIMP, the leading candidate
in this class is the lightest neutralino, a particle predicted
by the 
supersymmetric
extension of the Standard Model. We will discuss why neutralinos
are so interesting and will review the current and projected
experiments
for their detection.

\section{Dark matter candidates}

Since ordinary matter is baryonic, 
the most straightforward possibility is to assume also
this composition for the evasive dark matter. The contribution
from gas is not enough, 
so astrophysical bodies collectively known as MAssive Compact Halo 
Objects (MACHOs) are the main candidates.
This is the case of brown and white dwarfs, jupiter-like objects, 
neutron stars,
stellar black hole remnants. 
However, the scenario of Big-Bang nucleosynthesis, which
explains the origin of the elements after the Big Bang, 
taking into account measured abundances of helium, deuterium and
lithium sets a limit to the 
number of baryons that can exists in the Universe, 
namely 
%
\begin{equation}
\Omega_{\mbox{\tiny baryon}} \lsim 0.04 
\ .
\label{omegabaryon}
\end{equation}
This density is clearly small to account for the whole dark matter
in the Universe (see bounds (\ref{omegagalactic}),
(\ref{omegacluster}) and (\ref{omegaflow})).
The conclusion is that baryonic objects are likely
components of the dark matter 
but more candidates are needed.
This result is also confirmed by observations of MACHOs in our galactic 
halo
through their gravitational lensing effect 
on the light emitted by stars. Their contribution to the 
dark matter density is small.
Thus non-baryonic matter is required in the 
Universe. 

Particle physics provides this type of candidate for dark matter.
The three most promising are 
`axions', `neutrinos' and `neutralinos' with 
masses of the order of $10^{-5}$ eV, 30 eV and 100 GeV, 
respectively\footnote{The electronvolt (eV) is a unit of energy,
1 eV $\simeq$ 1.6$\times 10^{-19}$ J. Due to Einstein's equation, $E=mc^2$,
the masses of particles are
given in units of eV/$c^2\simeq 1.78\times 10^{-36}$ kg, usually shortened to
eV. For example, the mass of the electron is 0.51 MeV, where
the megaelectronvolt (MeV), 1 MeV=$10^{6}$ eV, is used.
The mass of the proton is of the order of 1 GeV, where
the gigaelectronvolt (GeV), 1 GeV = $10^{9}$ eV, is used.}. 
Although, as mentioned in Section~1, these particles are not
present in the current Standard Model of particle physics,
they are well motivated by
theories that attempt to unify the forces and particles of
Nature, i.e. by extensions of the Standard Model. 
Perhaps it is not a coincidence that such particles which may solve crucial
problems in particle physics, as we will discuss below, 
also solve the dark matter problem 
(it could be either a
big coincidence or a big hint). 

Before analyzing these candidates, let us mention that many others
have also been proposed.
Some of them are quite exotic and most of them
are ephemeral.

\subsection{Neutrinos}

The only dark matter candidates which are known to exist are
neutrinos. 
They are leptons, i.e. non-strongly-interacting elementary
particles, with zero charge and spin 1/2.
The 
Standard Model \cite{Kalmus} has three families or flavours of 
(left-handed) neutrinos $\nu_L$, each associated
with an electron-like lepton, i.e. there are electron neutrinos,
muon neutrinos and tau neutrinos.
These Standard Model
neutrinos are strictly massless because
there are no (right-handed) neutrinos $\nu_R$ that could combine with
$\nu_L$ to form a Dirac mass term through their interaction with
the Higgs doublet.

However, several extensions of the Standard Model 
do allow
neutrinos to have a mass. This is the case for example of grand unified
theories where the interactions of the Standard Model, i.e. the
strong interaction
$SU(3)$ and the electroweak interaction $SU(2)\times U(1)$ are combined 
as components of a single one, e.g. $SO(10)$ or $E_6$.
Then, $\nu_R$ appear in a natural way. 
Moreover, in recent years, observation of solar and
atmospheric 
neutrinos have indicated that one flavour might change to another.
Remarkably, this is a quantum process (neutrino oscillations) which
can only happen if the neutrino has a mass. The best evidence 
for neutrino mass comes
from the SuperKamiokande experiment in Japan
concerning atmospheric neutrino oscillations. The results of this experiment 
indicate a mass
difference of the order of 0.05 eV between the muon neutrino and the tau 
neutrino. If there
is a hierarchy among the neutrino masses 
(as it is actually the case not only for quarks where
e.g. the top mass is five orders of magnitude bigger than the up mass,
but also for  
electron-like leptons where
the electron mass is 0.51 MeV whereas
the muon mass is 105.6 MeV and the tau mass is
1777 MeV), then such a small mass difference implies that the neutrino masses
themselves lie well below 1 eV. This
is not cosmologically significant, as we will show below.  
On the other hand, there could be near mass degeneracy among the neutrino
families. In this case, if
neutrino masses $m_{\nu}\approx 30$ eV, they 
could still contribute significantly to the
non-baryonic dark matter in the Universe. 

These neutrinos left over from the Big Bang
were in thermal equilibrium in the early Universe
and decoupled when they were moving with relativistic velocities.
They fill the Universe in 
enormous quantities and their current number density
is similar to the one of photons
(by entropy conservation in the adiabatic expansion
of the Universe). In particular,  
$n_\nu=\frac{3}{11}\ n_\gamma$.
Moreover, the number density of photons 
is very accurately obtained from the cosmic microwave 
background measurements. The present temperature
$T\approx 2.725\ K$ implies
$n_{\gamma}\approx 410.5$ cm$^{-3}$. 
Thus one can compute the neutrino mass
density $\rho_\nu = m_{tot}\ n_\nu$, where 
$m_{tot}$
is basically the total mass due to all flavours of neutrino.
Hence,
\begin{equation}
\Omega_\nu \approx \frac{m_{tot}}{30\ \mathrm{eV}}\ .
\label{neutrinos}
\end{equation}
Clearly, neutrinos with $m_{\nu}\lsim 1$ eV cannot solve the
dark matter problem, but a neutrino with
$m_{\nu}\approx 30$ eV would give
$\Omega_{\nu}\approx 1$
solving it.
Unfortunately, due to the small energies involved,
detection of these cosmological neutrinos in the laboratory is
not possible. 



On the other hand, there is 
now significant evidence against neutrinos as the bulk of the
dark matter. 
Neutrinos belong to the so-called `hot' dark matter because
they were moving with relativistic velocities at the time the galaxies
started to form. But hot dark matter cannot reproduce correctly the
observed structure in the Universe.
A Universe dominated by neutrinos would form large structures first,
and the small structures later by fragmentation of the larger
objects. 
Such a Universe would produced a `top-down' cosmology, in which the
galaxies
form last and quite recently. 
This time scale seems incompatible with our
present ideas of galaxy evolution. 
This might be solved if galaxies were formed through topological defects
such as cosmic strings, but still is difficult to explain how neutrinos
could form the dark matter halos in dwarf galaxies.

This lead to fade away 
the initial enthusiasm for a
neutrino-dominated Universe. Hence, many cosmologists now favour an alternative
model, one in which the particles dominating the Universe are `cold'
(non-relativistic) rather than hot.
This is the case of the axions and neutralinos which we will study below.

\subsection{Axions}

The theory of the strong interaction, Quantum Chromodynamics (QCD), may
include in its Lagrangian the following invariant term formed
from the field strength $F_{\mu\nu}^a$\ , 
$\frac{\theta}{16\pi^2} tr (F_{\mu\nu} {\tilde F}^{\mu\nu})$,
where the parameter ($\theta/16\pi^2$) describes the strength of 
this term and 
${\tilde F}^{\mu\nu}=\frac{1}{2}\epsilon^{\mu\nu\rho\sigma}
F_{\rho\sigma}$.
In principle, this invariant can be reduced to a total divergence, 
hovewer, it cannot be dropped because non-perturbative effects in
QCD require such a term. The presence of this term
is potentially dangerous since it violates parity and CP,
and therefore there are important experimental bounds against it.
In particular, the stringent upper limit
on neutron dipole electric moment implies the bound $\theta<10^{-9}$.
This is the so-called strong CP problem. Why is this value
so small, when a strong interaction parameter would be expected
to be ${\cal O}(1)$? 

The most plausible solution to this naturalness problem 
is to show that $\theta$ is effectively
zero, and this was in fact the proposal made by Peccei and Quinn in 1977. 
They increased the number of Higgs bosons so that QCD together
with the electroweak theory has a global $U(1)$ symmetry.
Since the vacuum expectation values of the Higgses must be nonzero
to generate fermion masses, this symmetry is spontaneously broken
giving rise to a Goldstone boson, the so-called axion.
On the other hand, the CP violating phase becomes
a dynamical variable which is vanishing at the minimum of the
axion potential, as required.

In summary, axions are spin 0 particles with zero charge
associated with the spontaneous breaking of the
global $U(1)$ Peccei-Quinn symmetry, which was introduced
to dynamically solve the strong CP problem.
Although axions are 
massless at the classical level they pick up a small mass
by non-perturbative effects. The mass of the axion, $m_a$, and its 
couplings to ordinary matter, $g_a$, are proportional to $1/f_a$,
where $f_a$ is the 
(dimensionful) axion decay constant which is related to 
the scale of the symmetry breaking. 
In particular, the coupling of an axion with two fermions of
mass $m_f$, is given by $g_a\sim m_f/f_a$. Likewise, 
$m_a \sim \Lambda^2_{\mbox{\tiny QCD}}/f_a$, i.e. 
\begin{equation}
m_a 
\sim 10^{-5}~\mathrm{eV} \times \frac{10^{12}~\mathrm{GeV}}{f_a}\ .
\end{equation}

A lower bound
on $f_a$ can be obtained from the requirement that
axion emission does not over-cool stars. 
The supernova SN1987 put the strongest bound,
$f_a \gsim 10^9$ GeV. 
On the other hand, 
since coherent oscillations of the axion 
around the minimum of its potential may give an important contribution
to the energy density of the Universe, the requirement
$\Omega\lsim 1$ 
puts a lower bound on the axion mass implying $f_a \lsim 10^{12}$ GeV.
The combination of both constraints, astrophysical and cosmological,
give rise to the following window for the value of the axion constant
\begin{equation}
10^9~\mathrm{GeV}\lsim f_a \lsim 10^{12}~\mathrm{GeV}\ .
\end{equation}

The lower bound 
implies an extremely small coupling of the axion to 
ordinary matter and therefore a very large lifetime, larger than the
age of the Universe by many orders of magnitude. 
As a consequence, the axion is a candidate for dark matter.
Axions would have been produced copiously in the Big Bang, they
were never in thermal equilibrium and are always nonrelativistic
(i.e. they are cold dark matter).
In addition the upper bound implies that $m_a\sim 10^{-5}$ eV
if the axion is to be a significant component of the dark matter.

Since the axion can couple to two photons via fermion vacuum loops,
a possible way to detect it is through conversion to photon
in external magnetic field. Unfortunately, due to the small couplings
to matter
discussed above, we will not be able to produce axions in the
laboratory. On the other hand, relic axions are very abundant
(as we will show in Section~4, the density of dark matter particles
around the Earth is about 0.3 GeV\ cm$^{-3}$, since $m_a\sim 10^{-5}$
eV
there will be about $10^{13}$ axions per cubic centimeter)
and several experiments are trying already to detect axions or
are in project.
For example, an experiment at Lawrence Livermore National Laboratory 
(California, US) has reported in 1998 its first results
excluding a particular kind of axion of mass $2.9\times 10^{-6}$ eV
to 
$3.3\times 10^{-6}$ eV as the
dark matter in the halo of our galaxy.

\subsection{WIMPs}

As discussed in Section~1, weakly interacting massive particles,
the so-called WIMPs, are very interesting candidates for dark matter
in the Universe. They were in thermal equilibrium with the Standard
Model particles in the early Universe, and decoupled when
they were non-relativistic.
The process was the following.
When the temperature $T$ of the Universe was larger than the mass of the
WIMP, 
the number density of WIMPs and photons was roughly the same, 
$n_{\mbox{\tiny WIMP}}\propto T^3$,
and
the WIMP was annihilating with its own antiparticle into lighter
particles and vice versa.
However, shortly after the temperature dropped below the mass of the
WIMP,
$m$,
its number density dropped exponentially,
$n_{\mbox{\tiny WIMP}}\propto e^{-m
/T}$,
because only a small fraction of the light particles mentioned above
had sufficient kinetic energy to create WIMPs.
As a consequence, the WIMP annihilation rate dropped below the
expansion
rate of the Universe. At this point
WIMPs came away, they could not annihilate, and
their density is the same since then.  
Following these arguments,
the relic
density of WIMPs
can be computed with the result  
\begin{equation}
\Omega_{\mbox{\tiny WIMP}}  
\simeq \frac{7\times 10^{-27}\ \mathrm{cm^3\ s^{-1}}}{<\sigma_{ann}\ v >}\ , 
\label{relicdensity}
\end{equation}
where $\sigma_{ann}$ is the total cross section for annihilation of a
pair
of WIMPs into
Standard Model particles, $v$ is the relative velocity between the two WIMPs, 
$<..>$ denotes
thermal averaging, and the number in the numerator
is obtained using the value of the 
temperature of the cosmic background radiation,
the Newton's constant, etc. 
As expected from the above discussion about the early Universe, 
the relic WIMP density decreases with
increasing annihilation cross section.

Now we can understand easily why WIMPs are so good candidates for dark
matter.
If a new particle with weak interactions exists in Nature,
its cross section will be 
$\sigma\simeq \alpha^2/m_{\mbox{\tiny weak}}^2$,
where $\alpha\simeq {\cal O}(10^{-2})$ 
is the weak coupling and $m_{\mbox{\tiny weak}}\simeq {\cal O}(100$ GeV)
is
a mass of the order of the one of the W gauge boson, which is 
associated to the $SU(2)$ gauge group of
the Standard Model.
Thus 
one may obtain $\sigma\approx 10^{-9}$ GeV$^{-2}$.
Since at the freeze-out temperature $v$ is close to the speed of light,
one obtains\footnote{Here we are using natural units, where
1 GeV$^{-2}=0.389\times 10^{-27}$ cm$^2$.} 
$<\sigma_{ann}\ v>\approx 10^{-26}$ cm$^3$\ s$^{-1}$.
For our surprise, this number 
is remarkably close to the value that we need in eq.(\ref{relicdensity})
in order to obtain the observed density of the Universe.
This is a possible hint that new physics at the weak scale
provides us with a reliable solution to the dark matter problem.
This is a qualitative improvement with respect to the axion dark matter
case, where a small mass for the axion 
about $10^{-5}$ eV has to be postulated. 
As we will discuss in the next section, supersymmetry is a theory
that introduce new physics precisely at the weak scale, and
that predict a new particle, the neutralino, which could be stable and
therefore
the
sought-after dark matter.   

On the other hand, since WIMPs interact with ordinary matter with
roughly weak strength, their presence in galactic scales, and in particular
in our galaxy, raises
the hope of detecting relic WIMPs directly in a detector
by observing their
scattering on target nuclei through nuclear recoils.

\section{Neutralino dark matter}

Although supersymmetry was proposed 27 years ago, and 
despite the
absence of experimental verification,
it is still today
(together with string theory which anyway needs probably supersymmetry)
the most convincing theory in order to unify the physical laws \cite{Kane}.
Relevant theoretical arguments can be
given in its favour.

First of all, supersymmetry is a new 
type of symmetry since relates bosons and fermions. 
We know from the past
that symmetries are crucial in particle physics but in addition 
supersymmetry introduces a new
kind of unification between particles of different spin. 
In this sense, the Higgs scalar is
no longer a mysterious particle as it stands in the Standard Model, 
where it is introduced just in order to break the electroweak symmetry,
the
Supersymmetric Standard Model is naturally full of fundamental scalars 
(squarks,
sleptons and Higgs) related through supersymmetry 
with their fermionic partners
(quarks, leptons and Higgsino).
On the other hand, supersymmetry ensures the stability of the hierarchy
between the weak and the Planck scales. Since any fundamental theory
must contain gravity, due to quantum corrections
the masses of the scalar particles turn out to be 
proportional to the natural scale of the theory  
$M_{Planck}\simeq 1.2 \times 10^{19}~\mathrm{GeV}$. 
However, since these mass terms 
contribute to the Higgs potential, they should be
of the order of the 
electroweak scale (100-1000 GeV) in order to avoid any type 
of fine-tuning when breaking the electroweak symmetry. 
This problem of stabilizing the
scalar masses against quantum corrections is solved in supersymmetry 
since now the
scalar masses and the masses of their superpartners are related. 
As a consequence, the dangerous contributions of Standard Model
particles to quantum corrections 
are canceled with new ones which are present due to the
existence of the additional superpartners. 
Furthermore, the
joining of the three gauge coupling constants of the Standard Model 
at a single unification scale, in agreement with
experimental results from the LEP accelerator at 
CERN laboratory (Geneve, Switzerland), can only be obtained assuming
supersymmetry. 
Last, but not least,
the local version of supersymmetry leads to a partial unification
of the Standard Model with gravity, the so-called supergravity,
which is the low-energy limit of superstrings.

The price we have to pay for these marvellous properties, is the
introduction of a 
plethora of new particles. Not only the ones already mentioned 
(together with one more Higgs and Higgsino in order to cancel
anomalies),
but also
the spin-1/2 superpartners of the gauge bosons,
gluons, W's and B, i.e. the so-called gluinos, winos and bino.
All of them form the so-called Minimal Supersymmetric Standard Model,
the most simple supersymmetric extension of the Standard Model.
This is 
the most widely studied potentially realistic supersymmetric model.

Since ordinary particles and their superpartners only differ in spin,
supersymmetry would imply e.g. that the masses of gluinos are
vanishing as the masses of gluons, that the masses
of squarks and
sleptons are the same as the ones of quarks and leptons respectively,
and therefore there would be selectrons with a mass of 0.51 MeV, smuons
with a mass of 105.6 MeV, etc.
However 
there is no experimental 
evidence for such particles with those masses. 
Thus, in order for supersymmetry to play a realistic role in quark-lepton
physics, it must be a broken symmetry. 
Fortunately,
one can introduce consistently terms in the 
Lagrangian, the so-called `soft' terms, 
which explicitly break supersymmetry giving masses, for example,
to scalars. 
Moreover, these terms 
do not to spoil the supersymmetric
solution to the hierarchy problem (hence the name soft). 
It is important to remark that, as mentioned above, 
these mass terms 
should be
of the order of the 
electroweak scale.
Several accelerator experiments are in preparation 
in order to detect these predicted supersymmetric partners, e.g.
LHC at CERN will start operations in 2005 producing energies
about 1000 GeV.
It is worth recalling here that new physics at the electroweak scale
was the crucial assumption for WIMPs in order to obtain the observed density
of the Universe.

On the other hand, in the simplest supersymmetric model,
the Lagrangian contains interactions between one standard-model particle
and (always) two supersymmetric particles. For example, a quark-squark-gluino
interaction, which has in natural units 
the required dimension 4, is allowed. It is true that 
there are other possible interactions with dimension 4 and only one
supersymmetric particle, as e.g. quark-quark-squark or
squark-quark-lepton.
However they should be forbidden in order to avoid fast proton decay.
This yields important phenomenological implications.
Supersymmetric particles are produced or destroyed only in pairs
and therefore the lightest supersymmetric particle
is absolutely stable, implying that it might  
constitute a possible candidate for dark matter,
as first suggested by Goldberg in 1983. 
So supersymmetry, whose original motivation has nothing to do
with the dark matter problem, fulfils the two crucial requirements:
new physics at the electroweak scale with a stable particle.
In fact, since the lightest supersymmetric particle
is stable, another signature of major importance for this
supersymmetric model will be obtained in colliders.
One will be able to detect events with supersymmetric particle decays which
produce lots of missing energy (for instance, 
$e^+ e^- \rightarrow \mathrm{jet}+ \mathrm{ missing\ energy}$).



%
\begin{figure}[t]
\begin{center}
\epsfig{file= 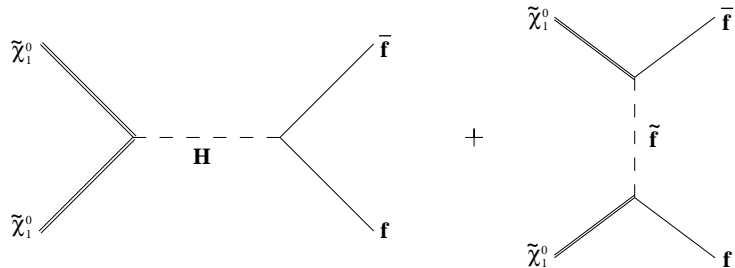,height=3.5cm,angle=0}
\end{center}
\caption{Feynman diagrams contributing to early-universe 
neutralino ($\tilde{\chi}^0_1$) annihilation into 
fermions through Higgses (H) and squarks and sleptons ($\tilde f$).}
\label{annihilation}
\end{figure}

It is remarkable that in most of the parameter
space of the Minimal Supersymmetric Standard Model
the lightest supersymmetric particle is an electrically neutral 
(also with no strong interactions) particle, called neutralino.
This is welcome since otherwise it 
would bind to nuclei and would be excluded as a candidate
for dark matter from unsuccessful searches for exotic heavy 
isotopes.

As a matter of fact there are four neutralinos, 
$\tilde{\chi}^0_i~(i=1,2,3,4)$, 
since they
are the physical 
superpositions of the fermionic partners of the neutral electroweak 
gauge bosons, 
bino and wino, and the
fermionic partners of the two  
neutral Higgs bosons, Higgsinos. 
Therefore the lightest neutralino, $\tilde{\chi}^0_1$, will be the 
dark matter candidate.
The experimental limit on its mass due to 
the negative searches at LEP is $m_{\tilde{\chi}^0_1} > 37$ GeV. 

Concerning the annihilation cross section contributing to the
density of the Universe in eq.(\ref{relicdensity}), 
there are numerous final states into which the 
neutralino can annihilate. The most important of these are the two body final 
states which occur at the tree level. Specially these are 
fermion-antifermion pairs, as shown in Fig.~\ref{annihilation}.
Many regions of the parameter space of the Minimal Supersymmetric
Standard Model produce values of the annihilation cross section 
in the interesting range mentioned below eq.(\ref{relicdensity}).
Therefore 
the neutralino is a very good candidate to account for the dark matter
in the Universe.

\section{The search for dark matter particles}

\subsection{Direct detection}

As discussed in Section~1, 
if neutralinos, or WIMPs in 
general, are the bulk of the dark matter,
they will form not only a background density in the Universe, but also
will cluster gravitationally with ordinary stars in the galactic halos.
In particular they will be present in our own galaxy, the Milky Way.
This raises the hope of detecting relic WIMPs directly, by experiments
on the Earth. In fact general studies of the possibility of dark matter
detection
began around 1982.
Since the detection will be in the Earth we need to know
the properties of our galaxy in order to be sure that
such a detection is feasible.


%
\begin{figure}[t]
\begin{center}
\begin{tabular}{c}
\epsfig{file= 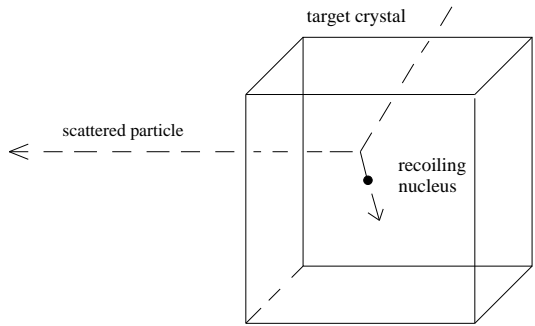,width=7cm,
}\\
\end{tabular}
\end{center} 
\vspace{0.5cm}
\caption{Elastic scattering of a dark matter particle with an
atomic nucleus in a detector.
} 
\label{scattering}
\end{figure}

As a matter of fact, rotation curves are much better known for external
galaxies than for ours, due to the position of the Earth inside the galaxy.
In any case analyses have been carried out with the conclusion
that indeed the Milky Way contains large amounts of dark 
matter \cite{ourgalaxy}.
Besides, 
some observational evidence seems to point at
a roughly spherical distribution of dark matter in the galaxy.
At the position of the Sun, around 8.5 kpc away from the galactic center,
the mean density of elementary particles trapped in the gravitational 
potential well of the galaxy is expected to be 
$\rho_{\mbox{\tiny dark}}\approx 5\times 10^{-24}$ gr\ cm$^{-3} \simeq
0.3$ GeV\ cm$^{-3}$. For WIMPs with masses about 100 GeV this means a number
density $n_{\mbox{\tiny WIMP}}\approx 3\times 10^{-3}$ cm$^{-3}$. 
In addition,
their velocity will be similar to the one of the Sun since they
move in the same gravitational potential well,
$v\approx 300$ km\ s$^{-1}$, implying a flux of dark matter particles
$J_{\mbox{\tiny WIMP}}
=n_{\mbox{\tiny WIMP}}\ v\approx 10^{5}$ cm$^{-2}$ s$^{-1}$
incident on the Earth.
Although this number is apparently large, the fact that WIMPs
interact weakly with matter makes the neutralino detection very 
difficult. Most of them will pass through matter without prevention.

\begin{figure}[t]
\begin{center}
\epsfig{file= 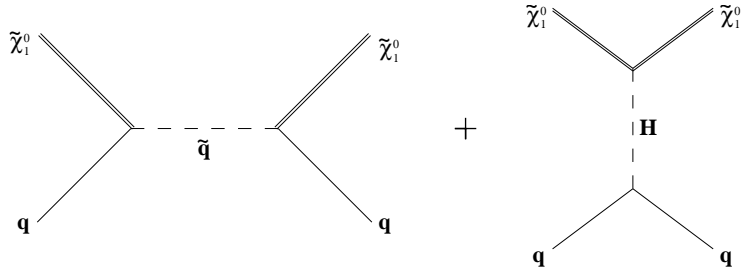,height=3.5cm,angle=0}
\end{center}
\caption{Feynman diagrams contributing to 
neutralino ($\tilde{\chi}^0_1$)--quark (q) cross section.
}
\label{crosssection}
\end{figure}

In any case, as suggested first by Goodman and Witten in 1985, 
direct experimental detection through elastic scattering with nuclei
in a detector, as shown schematically in Fig.~\ref{scattering},
is in principle possible.
A very rough estimate of the rate $R$ in a detector is the following.
A particle with a mass of the order of 100 GeV and electroweak interactions
will have a cross section 
$\sigma\approx 10^{-9}$ GeV$^{-2}$, as discussed in Subsection~2.3.
Thus for a material with nuclei composed of about 100 nucleons,
i.e. $m_{N}\sim 100$ GeV, one obtains
$R\sim J_{\mbox{\tiny WIMP}}\ \sigma/m_N \approx 10$ events\ kg$^{-1}$\ 
yr$^{-1}$.
This means that every day a few WIMPs, the precise number
depending on the number of kilograms of material, 
will hit an atomic nucleus in a detector.
Of course the above computation is just an estimate and one should
take into account in the exact computation the interactions of WIMPs
with quarks and gluons, the translation of these
into interactions with nucleons, and finally the translation of
the latter into interactions with
nuclei. In the case of neutralinos as WIMPs, diagrams
contributing to neutralino-quark cross section are shown
in Fig.~\ref{crosssection}. This scattering is elastic since
the nucleus recoils as a whole, giving billiard-ball collisions.
It is possible to check that many 
regions of the parameter space of the Minimal Supersymmetric
Standard Model produce values of the neutralino-nucleus cross section as those
mentioned above, and therefore giving rise to
a reasonable number of events ($10^{-5}$ to 10 events
per day per kilogram).

\subsection{Experimental techniques}

Experimentalists use three 
main techniques in order to
detect the nuclear recoil energy \cite{Smith}. The technique of 
measuring
ionization in solids, since the recoiling nucleus from a neutral
particle collision is itself charged and can also produce some
electron-hole excitation by collision with electrons.
The alternative technique of measuring 
ionization through the  
emission of photons in scintillation crystals. Finally, 
a small temperature rise.
For example when a 1 cm$^3$ of Silicon (Si) crystal is cooled to a very low
temperature ($\approx 120$ mK) in a dilution refrigerator, the heat
capacity ($\propto T^3$) is so low that even a few keV of deposited
energy raises the temperature by a measurable amount ($\sim 10^{-6}$ K).
This can be observed
since the recoiling nucleus 
(or more accurately the recoiling atom, since
at $v\sim 300$ km\ s$^{-1}$ most of the atomic electrons will remain bound to
the moving nucleus)
is stopped within $10^{-7}-10^{-6}$ cm
($\sim 10^{-14}$ s) releasing phonons.
Which method is better depends on the target material, and in fact
in many materials more than one of these effects may be observed.

On the other hand, for a 
slow moving ($\sim$ 300 km\ s$^{-1}$) and heavy
($\sim 100-1000$ GeV) particle forming the dark matter halo, 
the kinetic energy is very small, around
100 keV. The largest recoil energy transfer to a nucleus in the detector
is obtained when the mass of the WIMP and its mass are equal,
$m_{N}\sim 100$ GeV, but in any case it
will only be a few keV .   
Since cosmic rays with energies $\sim$ keV-MeV 
bombard the surface of the Earth, the experiments must have an
extremely good background discrimination.
In particular, neutrons coming from collisions between
cosmic-ray muons and nuclei produce nuclear recoils similar to
those expected from WIMPs  at a rate
$\sim$ 10$^{3}$ events kg$^{-1}$ day$^{-1}$.
Thus it is convenient to carry the experiments out in the deep underground,
in order to reduce the background 
by orders of magnitude 
in 
comparison with the sea level intensity.

In fact, this is still not enough since  
the detector has to be protected also 
against the natural radioactivity from the
surroundings (e.g. the rocks) and the materials of the detector itself.
This produces again neutrons but also X rays, gamma rays and beta rays giving
rise to electron recoils.
The latter may be a problem for detectors based only on
ionization or scintillation light since nuclear recoils with energies
of a few keV are much less efficient in ionizing or
giving light than electrons of the same energy.
Various protections aim to reduce these backgrounds. 
In particular, low radioactive materials, such as e.g.
high-purity copper or aged lead, are used for the shielding.
In addition, high-purity materials for the detector are also used.

\subsection{Experiments around the world}

Germanium is a very pure material and has been used for many years
for detecting `neutrinoless double beta decay'. Double beta decay
is a process
where two neutrons decay into protons, electrons and antineutrinos.
However if neutrinos are massive they cannot be produced.
These searches have reached extremely low background levels and
therefore
are very interesting for detecting dark matter particles.
In fact, $^{76}$Ge ionization detectors has been applied to WIMP searches
since 1987 \cite{Gedetectors}.
The best combination of data from these experiments, together with
the last data from the Heidelberg-Moscow 
\cite{HeidelbergMoscow} 
and IGEX experiments 
\cite{IGEX} 
located at
the Gran Sasso (L'Aquila, Italy) and Canfranc (Huesca, Spain) 
underground laboratories, respectively, 
have been able to exclude
a WIMP-nucleus cross section larger than $10^{-6}$ GeV$^{-2}$ 
for masses of WIMPs $\sim 100$ GeV, due to the negative search result.
Although this is a very interesting bound, it is still well above the
expected weak-interaction value $\sim 10^{-9}$ GeV$^{-2}$.

A generic problem of these type of detectors with germanium crystals,
given the small expected event rates,
is that only small masses ($\lsim$ 10 kg) 
can be assembled. However, this problem
disappears when sodium-iodide (NaI) scintillation detectors are
used. In addition, NaI gives good sensitivity over a wide range of
WIMP masses, since the mass of Na is small ($\sim$ 23 GeV) 
and the mass of I is large ($\sim$ 127 GeV).

Very intriguing results have been obtained in two 
experiments \cite{links} 
using
NaI detectors. The UK Dark Matter Collaboration (UKDMC) 
\cite{UKDMC}
has a scintillation detector
located 1100 metres below ground at the Boulby salt mine (Yorkshire, UK)
which comprises several NaI crystals ranging from
1 to 10 kg. 
It started taking data in 1997, and a number of
anomalous events have emerged in the data since the autumn of 1998.
Although spurious events due to several usual sources, as e.g. gamma rays,
neutrons, etc., have been investigated and dismissed,
a WIMP origin of the nuclear recoils is uncertain since
they were faster than the neutron calibrations used to simulate the 
WIMP interactions. Other (atypical) sources of the events are currently under
investigation.

\begin{figure}[t]
\begin{center}
\begin{tabular}{c}
\epsfig{file= 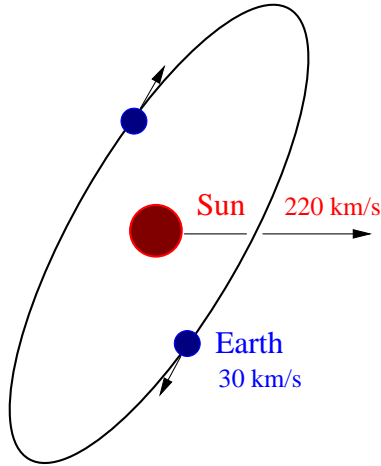,width=5cm
}\\
\end{tabular}
\end{center} 
\caption{
Earth's motion around the Sun.}
\label{sun} 
\end{figure}

A different method for discriminating a dark matter signal from
background is the one used by the DAMA collaboration, the so-called
annual modulation signature. 
As it is shown schematically 
in Fig.~\ref{sun}, as the Sun orbits the galaxy with velocity
$\approx 220$ km\ s$^{-1}$, the Earth orbits the Sun with
velocity $\approx 30$ km\ s$^{-1}$ and with the orbital plane
inclined at an angle of $60^{\circ}$ to the galactic plane.
As a consequence the dark matter flux will be larger in June
(when the Earth's rotation velocity adds to the Sun's velocity through the
halo) than in December
(when the two velocities are in opposite directions).
This fluctuation produces a rate variation $\approx 7\%$
between the two extreme conditions.
The DAMA experiment 
\cite{DAMA} 
investigates the annual modulation
of this signature. It 
involves physicists from Italy and China, and consists of 
about 100 kg of material (nine
9.70 kg NaI crystal scintillators to be precise) located
at the Gran Sasso laboratory built 1400 metres below ground.
Remarkably, they found that 
their detectors flashed more times in June than in December.
The data collected over four yearly cycles, since November 1996,
strongly favour  
the presence of a yearly modulation.
Besides, this signal is consistent
with a WIMP with mass
$\approx 30-200$ GeV 
and a
WIMP-nucleus cross section $\sigma\approx 10^{-6}-10^{-7}$ GeV$^{-2}$,
corresponding at a rate of about 1 event per kg per day.
DAMA group is confident about this result since they claim
to have ruled out other possible explanations, as e.g. temperature
changes.
It is worth noticing that 
although the above values for the cross section are still above the
expected weak-interaction value, they can be obtain in some
regions of the parameter space of supersymmetric models with
neutralino dark matter.

However the DAMA result is controversial, mainly because 
the negative search result obtained by the Cryogenic Dark Matter Search
(CDMS) experiment in the US 
\cite{CDMS}.
This is located just 10 metres below ground at Stanford University
in California,
and therefore it must discriminate 
WIMPs signals against interactions of background particles.
Two of the three detection techniques described above are used for
this discrimination, both the ionization and the temperature rise
produced during a recoil are measured.
This allows to discriminate electron recoils 
caused by interactions of background particles from
WIMP-induced nuclear recoils.
The ratio of deposited energies heat/ionization would be $\sim 2-3$
for the former and larger than 10 for the latter.
However, although neutrons are moderated by a 
25-cm thickness of polyethylene between the outer lead shield and
cryostat, an unknown fraction of them still survives.
Two data sets are used in this analysis: one consisting of 33 live days
taken with a 100-g Si detector between April and July 1998,
and another consisting of 96 live days taken with three 165-g Ge detectors
between November 1998 and September 1999.
Although four nuclear recoils are observed in the Si data set, they
cannot
be due to WIMPs, they are due in fact to the unvetoed neutron
background.
On the other hand, in the Ge detector thirteen nuclear recoils are 
observed per 10.6 kg per day, which is a similar rate to that expected
from the WIMP signal claimed by DAMA.
However, the CDMS group concludes that they are also due to neutrons.
These data exclude much of the region allowed by the DAMA annual
modulation signal.

\section{Future dark matter searches}

Due to this controversy
between DAMA and CDMS experimental results,
we cannot be sure whether or not the first direct evidence for the existence
of dark matter has already been observed.
Fortunately, the complete DAMA region will be tested by current dark matter
detectors.
This is for example the case of 
the IGEX experiment \cite{IGEX} mentioned above,
which is 
located at the Canfranc Tunnel Astroparticle Laboratory (2450 metres of 
water equivalent (m.w.e.)),
with an additional 1 kg-year of
exposure,
i.e. a few months of operation with two upgraded IGEX detectors.
Likewise the Heidelberg Dark Matter Search (HDMS) experiment 
\cite{HDMS}, 
which operates two ionization Ge detectors in
a unique configuration, was installed
at Gran Sasso in August 2000 and will also be able to test the DAMA
region
in the future.

In addition, DAMA and CDMS collaborations plan to expand their
experiments.
The DAMA collaboration will increase the amount of NaI crystals in 
its detector from 100 to 250 kg, 
which will make the 
experiment more sensitive to the annual modulation signal. 
On the other hand, the CDMS collaboration
is planning to move its detector 
to the abandoned Soudan mine in Minnesota (approximately 700 metres
below ground), increasing also  
the mass of its 
Ge/Si targets to 10 kg.
Of course this will clarify completely whether or not the thirteen events
found
by the current experiment at Stanford are due to neutrons.
This experiment will be able to test a WIMP-nucleus cross section
$\sigma > 10^{-9}$ GeV$^{-2}$

In the light of the polemical results from the DAMA and CDMS
collaborations, a new generation of very sensitive 
experiments have been proposed
all over the world. For instance only in the Gran Sasso laboratory 
there will be five experiments searching for WIMPs.
Apart from the two already discussed DAMA and HDMS, 
there are three other experiments in prospect, 
CRESST, CUORE and GENIUS. 
For example, 
the Cryogenic Rare Event Search using Superconducting Thermometers (CRESST)
experiment \cite{CRESST}, 
which
involves the Max Planck Institute for Physics, 
the Technical University in Munich, the Gran Sasso, and
the University of Oxford, measures simultaneously phonons and
scintillation light distinguishing the nuclear recoils
from the electron recoils cause by background radioactivity.
In contrast to other experiments, CRESST detectors allow the
employment of a large variety of target materials,
such as e.g. sapphire or tungsten. This allows a better sensitivity
for detecting the WIMPs.
In the long term the present CRESST set-up permits
the installation of a detector mass up to 100 kg, 
which will test a WIMP-nucleus cross section slightly smaller than 
the CDMS Soudan one discussed above.


But
the most sensitive 
detector will be the GErmanium in liquid NItrogen Underground Setup
(GENIUS) \cite{HDMS}, 
which will be able to test a WIMP-nucleus cross section
as low as $\sigma\approx 10^{-10}$ GeV$^{-2}$. 
Indeed such a sensitivity covers almost the full range
of the parameter space of supersymmetric models with
neutralinos as dark matter.
The GENIUS project 
is based on the idea to operate an array of 100 kg of Ge crystals directly in 
liquid nitrogen.
The latter,
which is very clean with respect to radiopurity, can act 
simultaneously 
as cooling medium and shield against external activities,
using a sufficiently large tank $\sim$ 12 metres in diameter at least.
It has been shown using Monte Carlo simulations
that with this 
approach the unwanted background is reduced by three to four orders 
of magnitude. 
In order to demonstrate the feasibility of the GENIUS project
a GENIUS Test-Facility (GENIUS-TF)
has been approved \cite{HDMS}.

In addition to these Gran Sasso experiments, 
there is also the French experiment EDELWEISS which is being carried out
at Modena underground laboratory, 
the French project MACHe3 at the Joseph Fourier University,
the PICASSO project at the University of
Montreal, the ORPHEUS project at the University of Bern,
and the ZEPLIN project at the Boulby mine.
For example, the latter is an experiment carried out between 
the UKDMC collaboration, the University of California at Los Angeles,
the Institute of Theoretical and Experimental Physics in Moscow,
CERN and the University of Torino in Italy.
It consists of a series of xenon detectors where the
nuclear recoil produces both an ionization and a scintillation signal,
giving a discrimination power 10--100 times better than for NaI.

Efforts to build detectors sensitive to the directional
dependence, i.e. recoil away from direction of Earth motion,
are also being carried out.
This is an extension of the idea of annual modulation.
To reconstruct such a three-dimensional direction is not
simple, but UKDMC collaboration together
with Temple University in Philadelphia, Occidental College in Los
Angeles and the University of California at San Diego,
are studying the possibility of using 
an ionization xenon-gas detector, the so-called
DRIFT, for this issue. 
The arrival time of the ionization signal will be used to
reconstruct the event in three dimensions.

Let us finally mention 
that there are other promising methods for the (indirect) detection 
of WIMPs in the halo \cite{Wilczek}. 
For example WIMPs passing through the Sun and/or Earth 
may be slowed below escape velocity by elastic scattering.
Thus they will accumulate and annihilate in the center
producing neutrinos. These can be detected in underground
experiments, specially through the muons produced 
by their interactions in the rock.
Another way of detecting WIMPs indirectly is through anomalous
cosmic rays produced by their annihilations in the galactic halo.

\section{Conclusions}

Nowadays there is overwhelming evidence that most of the mass in the universe
(90\% and probably more) 
is some (unknown) non-luminous `dark matter'.
At galactic and cosmological scales it 
only manifests through 
its gravitational interactions with 
ordinary matter.
However, at microscopical scales it might manifest through weak interactions,
and this raises the hope 
that it may be detected in low-energy particle physics
experiments.

One of the most interesting candidates for dark matter is
the so-called neutralino, a particle predicted by the
supersymmetric extension of the Standard Model.
These neutralinos are stable and therefore may be left over from the
Big
Bang.
Thus they 
will cluster gravitationally with ordinary stars in the galactic halos, 
and 
in particular they will be present in our own galaxy, the Milky Way.
As a consequence there will be a flux of these dark matter particles
on the Earth. 

Many experiments have been carried out around the world in order to detect this
flux. One of them even claims to have detected it.
Unfortunately, this result is controversial because of the negative
search result obtained by another experiment in the same range of
parameters.
Thus we will have to wait for the next generation of experiments,
which are already starting operations or in project, to 
be sure
whether or not the neutralino, or generically, a weakly interacting
massive
particle, is the evasive dark matter filling the whole Universe.

In summary, underground physics as the one discussed here is crucial 
in order
to detect dark matter. Even if neutralinos are discovered
at future particle accelerators such as LHC, only their direct detection
due to their presence in our galactic halo will confirm that they
are the sought-after dark matter of the Universe.
\\

\noindent {\bf Acknowledgments}
\vspace{0.4cm}

\noindent
We gratefully acknowledge D.G. Cerde\~no for his valuable help


 
 



\begin{thebibliography}{99}
\bibitem{Smith} A complete 
analysis
of different ideas proposed for dark matter detection
is available from the review mentioned in ref.~\cite{ourgalaxy}.
\end{thebibliography}
\end{document}